\documentstyle[aps,prl,twocolumn]{revtex}
\input{epsf}
\begin{document}
\draft
\title{Quantum Nonlocality in Phase Space}
\author{Konrad Banaszek and Krzysztof W\'{o}dkiewicz\cite{unm}}
\address{Instytut Fizyki Teoretycznej, Uniwersytet Warszawski,
Ho\.{z}a~69, PL--00--681~Warszawa, Poland}
\date{\today}
\maketitle

\begin{abstract}
We propose an experiment demonstrating the nonlocality of a quantum
singlet-like state generated from a single photon incident on a beam
splitter. Each of the two spatially separated apparatuses in the
setup performs a
strongly unbalanced homodyning, employing a single photon
counting detector. We show that the correlation functions violating the
Bell inequalities in the proposed experiment are given by the
joint two-mode $Q$-function and the Wigner function of the optical
singlet-like state. This establishes a direct relationship between two
intriguing aspects of quantum mechanics: the nonlocality of entangled
states and the noncommutativity of quantum observables, which underlies the
nonclassical structure of phase space quasidistribution functions.
\end{abstract}
\pacs{PACS Number(s): 03.65.Bz, 42.50.Dv}

A fundamental step  providing a bridge between classical and quantum
physics has been given  by Wigner in form of a quantum mechanical phase
space distribution: the Wigner function \cite{WignPR32}.
From the pioneering work of Weyl, Wigner and Moyal, it follows
that the noncommutativity of quantum observables leads to a real
abundance of different in form quantum mechanical phase space
quasidistributions.  A description of quantum phenomena in terms of the
Wigner or the positive-$Q$ quasidistributions, provided a milestone step
towards a $c$-number formulation of quantum effects in phase space
\cite{PhaseSpaceReview}.

Due to  Einstein, Podolsky, and Rosen (EPR) \cite{EPR}, followed  by
the seminal contribution of Bell \cite{Bell65}, the meaning of quantum
reality and quantum nonlocality has become a central issue of the
modern  interpretation and understanding of quantum phenomena
\cite{peres}.  Such concepts like entanglement and quantum nonlocality
have generated a real flood of  theoretical work devoted to various
connections of the quantum description  with different views or
representations of the quantum formalism. 

Despite all these theoretical works  a direct link between various
phase space distributions and the nonlocality of quantum mechanics has
been missing.  In some works \cite{EPRandEPW} the quantum phase space
has been treated as a model for a hidden variable theory, and the
incompatibility of quantum mechanics with local theories has been
attributed to the nonpositive character of the Wigner function.  In
this context it has been argued that the original EPR wave function
cannot violate the position-momentum Bell inequality, because the
corresponding Wigner function is positive everywhere.

It is the purpose of this Letter to propose an experimental
demonstration of nonlocal effects in phase space exhibited by a quantum
optical singlet-like state generated from a single photon. The
entanglement  will be represented by a correlated state of light, which
refers to two spatially separated modes of the electromagnetic field.
We show that the proposed experiment establishes a direct relationship
between  quantum nonlocality and the positive phase space $Q$-function,
as well as the nonpositive Wigner function. We demonstrate that for a
certain class of experiments these two quasiprobability distributions
{\em are} nonlocal correlation functions violating Bell's
inequalities.

In this Letter we propose  a realistic photon counting experiment which
leads directly to a measurement that is described by the   phase space
$Q$-function or the Wigner function.  We show that these functions are
equal to observable joint photon count correlations and as such can be
put to test of local realism in form of  Bell's inequalities for an
entangled single photon.  Our approach is different from all the
previous discussions involving the relation of quantum nonlocality and
the phase space quasiprobability distributions.  To the best of our
knowledge, no such direct relation between various phase space
quasidistributions and the nonlocality of quantum correlations has ever
been satisfactorily established.

The link of quantum nonlocality to the $Q$-function is a rather
surprising result, since this particular distribution function is
positive everywhere, which has been considered as a loss of
quantum properties due to simultaneous measurement of canonically
conjugated observables.

\begin{figure}

\centerline{\epsfxsize=3in\epsffile{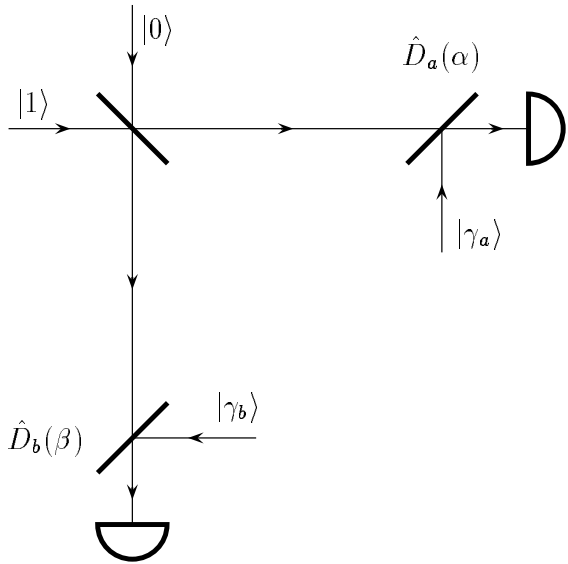}}

\vspace{1cm}

\caption{The optical setup proposed to demonstrate quantum
nonlocality in phase space. A single photon incident on a 50:50 beam
splitter generates a quantum singlet-like state.  The measuring devices
are photon counting detectors preceded by beam splitters. The beam
splitters have the transmission coefficient close to one, and strong
coherent states injected into the auxiliary ports. In this limit, they
effectively perform coherent displacements $\hat{D}_{a}(\alpha)$ and
$\hat{D}_{b}(\beta)$ on the two modes of the input field.}
\label{Fig:Setup}
\end{figure}

The setup to demonstrate quantum nonlocality in phase space is
presented in Fig.~\ref{Fig:Setup}.
A single photon impinges onto a 50:50 beam
splitter. The quantum state written in terms of the outgoing modes,
which we will label with $a$ and $b$, is of the form analogous to the
singlet state of two spin-$1/2$ particles \cite{SinglePhoton}:
\begin{equation}
|\Psi\rangle = \frac{1}{\sqrt{2}}(|1\rangle_a |0\rangle_b
- |0\rangle_a |1\rangle_b).
\end{equation}
Each of the measuring apparatuses consists of a photon counting
detector preceded by a beam splitter with the power transmission $T$.
The second input port of the beam splitter is fed with a highly excited
coherent state $|\gamma\rangle$. As it is known 
\cite{Homodyning}, in the limit
$T\rightarrow 1$ and $\gamma\rightarrow \infty$, the effect of the
beam splitter is described by the displacement operator
$\hat{D}(\sqrt{1-T}\gamma)$ with the parameter equal to the amplitude
of the reflected part of the coherent state.  In the following, we will
assume that this limit describes sufficiently well the measuring
apparatuses.

The first type of the measurement we will consider is the test for the
presence of photons. This is a more realistic case, as the
most efficient detectors available currently
for single-photon level light, namely the
avalanche photodiodes operating in the Geiger mode, are not capable of
resolving the number of photons that triggered the output signal.
This type of measurement is described by a pair of two orthogonal
projection operators depending on the coherent displacement
$\alpha=\sqrt{1-T}\gamma$:  
\begin{eqnarray} \hat{Q}(\alpha) & = &
\hat{D}(\alpha)|0\rangle\langle 0| \hat{D}^{\dagger}(\alpha) \nonumber
\\ \hat{P}(\alpha) & = & \hat{D}(\alpha)
\sum_{n=1}^{\infty}|n\rangle\langle n| \hat{D}^{\dagger}(\alpha)
\end{eqnarray} 
which satisfy the completeness relation:
\begin{equation}
\hat{Q}(\alpha) + \hat{P}(\alpha) = \hat{\openone}.
\end{equation}
In the following, we will use the indices $a$ and $b$ to
refer to the two apparatuses.

In contrast to the standard approach, we will be interested in events
when {\em no photons} were registered.
The joint probability of no-count events simultaneously in both the
detectors is:
\begin{eqnarray}
Q_{ab}(\alpha,\beta) & = & \langle\Psi| \hat{Q}_{a}(\alpha) \otimes
\hat{Q}_{b}(\beta) |\Psi\rangle \nonumber \\
& = & \frac{1}{2} |\alpha-\beta|^2
e^{-|\alpha|^2-|\beta|^2}
\end{eqnarray}
where $\alpha$ and $\beta$ are coherent displacements for the modes $a$
and $b$, respectively. The probabilities on single detectors are:
\begin{eqnarray}
Q_a(\alpha) & = & \langle \Psi | \hat{Q}_{a}(\alpha) 
\otimes \hat{\openone}_b
|\Psi\rangle
= \frac{1}{2}(|\alpha|^2+1)e^{-|\alpha|^2}, \nonumber \\
Q_b(\beta) & = & \langle \Psi | 
\hat{\openone}_{a} \otimes
\hat{Q}_{b}(\beta) |\Psi\rangle
= \frac{1}{2}(|\beta|^2+1)e^{-|\beta|^2}.
\end{eqnarray}

\begin{figure}

\centerline{\epsfxsize=3.375in\epsffile{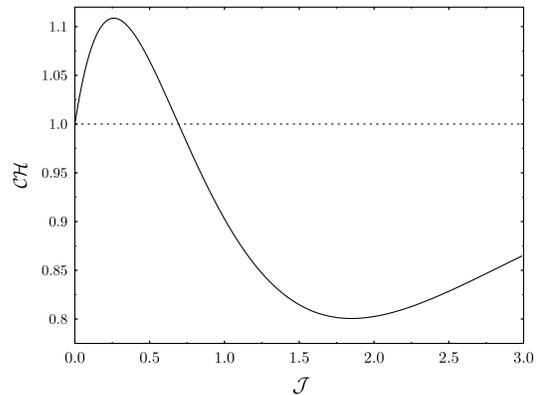}}

\caption{The plot of the Clauser-Horne combination defined in 
Eq.~(\protect\ref{Eq:CHdef}) as a function of the intensity
of coherent displacements ${\cal J}= |\alpha|^2 = |\beta|^2$,
for opposite phases $\beta=-\alpha$. The dotted line indicates
the upper bound imposed by local theories.}
\label{Fig:CHplot}
\end{figure}

The measurement is now performed for two settings of the coherent
displacement in each of the apparatuses: either zero, or $\alpha$
for the mode $a$ and $\beta$ for the mode $b$. From the resulting
four different correlation functions we build the Clauser-Horne
combination \cite{ClauHornPRD74}:
\begin{eqnarray}
{\cal CH} & = & Q_a(0) + Q_b(0) - Q_{ab}(0,0) \nonumber \\
\label{Eq:CHdef}
& & - Q_{ab}(\alpha,0) - Q_{ab}(0,\beta) + Q_{ab}(\alpha,\beta),
\end{eqnarray}
which for local theories satisfies the inequality
$0 \le {\cal CH} \le 1$. We will take the coherent displacements
to have equal magnitudes $|\alpha|^2 = |\beta|^2 = {\cal J}$ and 
an arbitrary phase difference $\beta = e^{2i\varphi}\alpha$. For
these values we obtain
\begin{equation}
\label{Eq:CHviolation}
{\cal CH} = 1 - {\cal J}e^{-{\cal J}} + 2 {\cal J} e^{-2 {\cal J}}
\sin^2 \varphi.
\end{equation}
As depicted in Fig.~\ref{Fig:CHplot}, this result violates the upper bound
imposed by local theories. The violation is most significant for the
phase $\varphi$ which maximizes the last term in
Eq.~(\ref{Eq:CHviolation}). This takes place when the coherent
displacements have opposite phases $\beta=-\alpha$.

Let us note that when no displacements are applied, the detectors
measure the bare state $|\Psi\rangle$. This state contains a single
photon in the sense that it is an eigenstate of the total photon number
operator $\hat{n}_{a}+\hat{n}_{b}$ with an eigenvalue 1. In this case
$Q_{ab}(0,0)=0$, which means that the photon is always registered by
one of the detectors.

The only measurement that is required to demonstrate the nonlocality of
this state requires single and joint registration of {\em no photons}.
When the state is not shifted, this measurement is described by the
projection on the vacuum state $|0\rangle$. Furthermore, application of
a coherent displacement $\hat{D}(\alpha)$ is equivalent to the
projection on a coherent state $|\alpha\rangle$.  And here comes the
most striking link of the quantum nonlocality with the phase space
quasidistribution. $Q_{ab}(\alpha,\beta)$ is consequently equal, up to a
constant $1/\pi^2$, to the joint $Q$-function of the state
$|\Psi\rangle$. The operator $\hat{Q}(\alpha)$, defined above, represents
a projection on a coherent state $|\alpha\rangle$, and the correlation
function is:
\begin{equation}
Q_{ab}(\alpha,\beta) = |\langle \alpha,\beta|\Psi\rangle|^2,
\end{equation}
where $|\alpha,\beta\rangle = |\alpha\rangle_a \otimes |\beta\rangle_b$.
The probabilities of no-count events on single detectors are given by
marginal $Q$-functions:
\begin{eqnarray}
Q_a(\alpha) & = &
\langle\alpha| \text{Tr}_b (|\Psi\rangle\langle\Psi|)
|\alpha\rangle_{a}, \nonumber \\
Q_b(\beta) & = &
\langle\beta| \text{Tr}_a (|\Psi\rangle\langle\Psi|)
|\beta\rangle_{b}.
\end{eqnarray}
Thus, we now clearly see that the $Q$-function contains direct
information on nonlocal quantum correlations. If a four-point
combination of the type given in Eq.~(\ref{Eq:CHdef}) violates the
inequality $0 \le {\cal CH} \le 1$, this immediately certifies the
nonlocal properties of the quantum state. This definition has an obvious
operational meaning, as we have discussed an experiment in which
the nonlocal character of the $Q$-function can be readily tested. 

Following a more traditional approach,
the combination ${\cal CH}$ defined in Eq.~(\ref{Eq:CHdef})
can be also related to the probabilities 
of registering photons by the detectors. A simple calculation shows 
that ${\cal CH}$ can also be expressed as:
\begin{eqnarray}
{\cal CH} & = & P_a(0) + P_b(0) - P_{ab}(0,0) \nonumber \\
& & - P_{ab}(\alpha,0)
-P_{ab}(0,\beta) + P_{ab}(\alpha,\beta)
\end{eqnarray}
where $P_a(\alpha)$, $P_b(\beta)$, and $P_{ab}(\alpha,\beta)$
are given by the expectation values over the state $|\Psi\rangle$
of the operators: $\hat{P}_a(\alpha)$, $\hat{P}_b(\beta)$,
and $\hat{P}_a(\alpha)\otimes \hat{P}_b(\beta)$, respectively.

In order to give an operational meaning to the Wigner function,
we will now consider the case when the detectors are capable of
resolving the number of absorbed photons. Let us assign to each
event $+1$ or $-1$, depending on whether an even or an odd number
of photons has been registered. This measurement is described
by a pair of projection operators:
\begin{eqnarray}
\hat{\Pi}^{(+)} (\alpha) & = & \hat{D}(\alpha) \sum_{k=0}^{\infty}
|2 k \rangle \langle 2k|
\hat{D}^{\dagger}(\alpha) \\
\hat{\Pi}^{(-)} (\alpha) & = & \hat{D}(\alpha) \sum_{k=0}^{\infty}
|2 k+1 \rangle \langle 2k+1 |
\hat{D}^{\dagger}(\alpha).
\end{eqnarray}
Using these projections, we construct the correlation function between
the outcomes of the apparatuses $a$ and $b$.
This correlation has a clear analogy to spin or to photon
polarization joint measurements and it is given by the expectation
value of the operator:
\begin{equation}
\hat{\Pi}_{ab}(\alpha,\beta) =   
(\hat{\Pi}_a^{(+)}(\alpha) - \hat{\Pi}_a^{(-)}(\alpha))
  \otimes
(\hat{\Pi}_b^{(+)}(\beta) - \hat{\Pi}_b^{(-)}(\beta)).
\end{equation}
This quantity is proportional to the joint two-mode Wigner function
of the state $|\Psi\rangle$. This link becomes obvious if we rewrite
$\hat{\Pi}_{ab}(\alpha,\beta)$ to the form:
\begin{equation}
\hat{\Pi}_{ab}(\alpha,\beta) = \hat{D}_{a}(\alpha) 
\hat{D}_{b}(\beta) (-1)^{\hat{n}_a + \hat{n}_b}
\hat{D}_{a}^{\dagger}(\alpha)
\hat{D}_{b}^{\dagger}(\beta)
\end{equation}
showing that the correlation function is given by the displaced 
parity operator $(-1)^{\hat{n}_a+\hat{n}_b}$, which is one of 
equivalent definitions of the Wigner function \cite{WignerParity}.
It is a striking result, that the nonlocality in a dichotomous
correlation measurement in our setup is given directly by the phase
space Wigner function of the state $|\Psi\rangle$.

\begin{figure}

\centerline{\epsfxsize=3.375in\epsffile{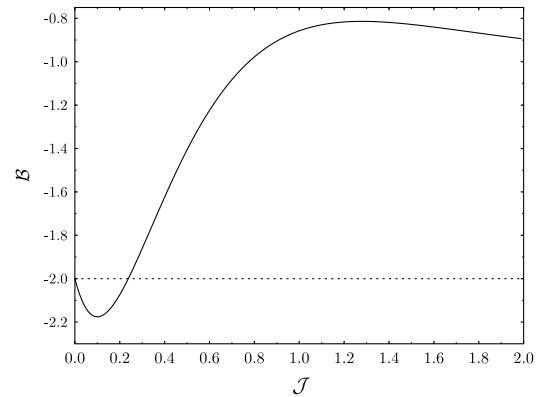}}

\caption{The plot of the combination defined in
Eq.~(\protect\ref{Eq:Bdef}) as a function of the magnitude
of coherent displacements parameterized with
${\cal J} = |\alpha|^2 = |\beta|^2$, for $\beta=-\alpha$.
The dotted line indicates the lower bound imposed by local theories.}
\label{Fig:Bplot}
\end{figure}

An easy calculation yields the expectation value of the operator
$\hat{\Pi}_{ab}(\alpha,\beta)$ over the state $|\Psi\rangle$:
\begin{eqnarray}
\Pi_{ab}(\alpha,\beta) & = & \langle\Psi| \hat{\Pi}_{ab}(\alpha,
\beta) | \Psi\rangle \nonumber \\
& = & (2|\alpha-\beta|^2 - 1) e^{-2|\alpha|^2-2|\beta|^2}.
\end{eqnarray}
Now we consider the combination \cite{CHSH}:
\begin{equation}
\label{Eq:Bdef}
{\cal B} = \Pi_{ab}(0,0) + \Pi_{ab}(\alpha,0)
+ \Pi_{ab}(0,\beta) - \Pi_{ab}(\alpha,\beta)
\end{equation}
for which local theories impose the bound
$-2\le {\cal B} \le 2$. Again we will take equal magnitudes of
the coherent displacements
$|\alpha|^2 = |\beta|^2 = {\cal J}$ and 
 a certain phase difference between them 
$\beta=e^{2i\varphi}\alpha$. Then the combination ${\cal B}$ takes 
the form:
\begin{equation}
{\cal B} = -1 +(4{\cal J}-2)e^{-2{\cal J}} - (8{\cal J}
\sin^2 \varphi -1)e^{-4{\cal J}},
\end{equation}
which, as shown in Fig.~\ref{Fig:Bplot}, for sufficiently small
intensities ${\cal J}$ violates the lower bound of the inequality
imposed by local theories.  As before, the strongest violation is
obtained for $\varphi=\pi/2$, i.e., when the coherent displacements
have opposite phases.

It is now an interesting question whether the nonlocality of the Wigner
function exhibited in the proposed experiment is connected to its
nonpositivity.
The Wigner function of the state $|\Psi\rangle$, containing only one photon,
is not
positive and exhibits the nonlocal character of quantum correlations.
The nonlocal character of this phase space function is directly
measured in an experiment involving a detection that resolves the
number of absorbed photons. However, it should be pointed out that  the
above  measurement for an incoherent  mixtures of the two components
forming the state $|\Psi\rangle$ leads to a joint  correlation equal
to $(2|\alpha|^2+2|\beta|^2 - 1) e^{-2|\alpha|^2-2|\beta|^2}$. Note
that this joint correlation is the Wigner function of the incoherent
mixture. This function is nonpositive, but it does not exhibit any
quantum interference effects and as a result  the Bell inequality is
not violated in this case. This shows, that the nonpositivity of the
Wigner function does not automatically guarantee violation of local
realism \cite{AnotherPaper}.

In conclusion, we have demonstrated that phase space quasidistribution
functions: the Wigner function and the $Q$-function carry explicit
information on nonlocality of entangled quantum states. This is due to the
fact that these two quasidistributions directly correspond to nonlocal
correlation functions which can be measured in a class of photon
counting experiments involving application of coherent displacements. 

{\it Acknowledgements.} This research was partially supported by
the Polish KBN grants and by Stypendium Krajowe dla M{\l}odych
Naukowc\'{o}w Fundacji na rzecz Nauki Polskiej.

\begin{references}
\bibitem{AnotherPaper}
This and other issues related to the nonpositivity and the nonlocality
of the Wigner function will be addressed in a separate publication.
\end{references}
\end{document}